\documentclass[12pt]{article}
\usepackage{graphicx}
\usepackage{lineno}
\usepackage{amsmath}
\newcommand\pubnumber{SNSN-323-63}
\newcommand\pubdate{\today}

\def\institute{Physikalisches Institut\\
Rheinische Friedrich-Wilhelms-Universit\"{a}t Bonn, GERMANY}
\def\support{\footnote{Speaker}}

\def\Title#1{\begin{center} {\Large #1 } \end{center}}
\def\Author#1{\begin{center}{ \sc #1} \end{center}}
\def\Address#1{\begin{center}{ \it #1} \end{center}}

\newcommand\pubblock{\rightline{\begin{tabular}{l} \pubnumber\\
         \pubdate  \end{tabular}}}
\newenvironment{Abstract}{\begin{quotation}  }{\end{quotation}}
\newenvironment{Presented}{\begin{quotation} \begin{center} 
             PRESENTED AT\end{center}\bigskip 
      \begin{center}\begin{large}}{\end{large}\end{center} \end{quotation}}

\begin{document}
\begin{titlepage}
	
\pubblock

\vfill
\Title{Recent measurements of associated single top-quark production cross-section with the ATLAS detector}
\vfill
\Author{ Irina A. Cioar\u{a}\support  \, on behalf of the ATLAS Collaboration}
\Address{\institute}
\vfill
\begin{Abstract}

%A measurements of the inclusive ccr$Wt$ production were published by the ATLAS and CMS collaborations, using 2012 data to search for events in which both the $W$ boson and top quark decays include an electron or a muon. The cross-section for the associated production of a single top quark has roughly tripled with the increase in centre-of-mass energy from 8 to 13 TeV, making it possible to perform measurements in the $Wt$-channel with the first available LHC datasets at this energy.\\
The measurement of the inclusive $Wt$ cross-section at 13 TeV is performed using 3.2$\,\text{fb}{^{-1}}$ of proton--proton collision data collected by the ATLAS detector at the LHC in 2015. Events are required to have at least one jet and two opposite sign leptons. The cross section is measured to be $\sigma_{Wt} = 94\pm 10\text{(stat.)}^{+28}_{-22}\text{(syst.)}\pm 2\text{(lumi})$pb, corresponding to an observed significance of 4.5\,$\sigma$ (3.9\,$\sigma$ expected). The result is consistent with the Standard Model prediction.

\end{Abstract}
\vfill
\begin{Presented}
$9^{th}$ International Workshop on Top Quark Physics\\
Olomouc, Czech Republic,  September 19--23, 2016
\end{Presented}
\vfill
\end{titlepage}
\def\thefootnote{\fnsymbol{footnote}}
\setcounter{footnote}{0}

\section{Introduction}
Single top quark production occurs via the electroweak interaction, through one of the three possible mechanisms: $s$-, $t$- and $Wt$ channel. These are excellent probes of the $Wtb$ coupling. 

 At 13 TeV the associated production of a top quark and a $W$ boson is the second largest one in terms of cross-section.

First observations of $Wt$ associated production were published by the ATLAS  \!\cite{Aad:2015eto} and CMS \!\cite{Chatrchyan:2014tua} collaborations, using the full Run~1 dataset to search for events in which both the $W$ boson and top quark decays include an electron or a muon in the final state. The cross-section for the associated production of a single top quark has roughly tripled with the increase in centre-of-mass energy from 8 to 13 TeV, making it possible to perform measurements in the $Wt$-channel with the first available LHC datasets at this energy.

These proceedings present a first measurement of the $Wt$ dilepton channel cross-section at 13 TeV \!\cite{Aaboud:2016lpj}, using 3.2fb$^{-1}$ of proton--proton collision data collected by the ATLAS detector  \!\cite{Aad:2008zzm} in 2015.

\section{Analysis strategy}

$Wt$ events that match the dilepton final-state topology are selected and sorted into different data sets. Exactly two opposite sign isolated leptons\footnote{Leptons refers to only electrons and/or muons throughout the text.} with $p_T ^{\text{lepton}}> 20 \text{ GeV}$ are required. Additionally, one of them must have a transverse momentum higher than $25 \text{ GeV}$ and at least one of the two has to be triggered on.  
All jets must have $p_T^{\text{jets}} > 25 \text{ GeV}$. Two signal regions are defined based on the number of jets. In both cases, exactly one jet identified as a b-jet is required (called 1j1b) or two jets, one of which must be $b$-tagged (2j1b). In order to control the main background coming from top quark pair production, a dedicated control region with two jets that are both $b$-tagged is defined. The agreement between data and Monte Carlo is controlled in two distinct regions that must include either 1 or 2 jets and no requirement on the number of $b$-tagged jets (1j0b and 2j0b). 

%
%\begin{figure}[htb]
%	\centering
%	\includegraphics[height=1.7in]{cuts}
%	\caption{bla}
%	\label{tab:cuts}
%\end{figure}

%%%%%%%%%%%%%%%%%%%%%%%%%%%%%%%%%%%%%%%%%%%%%%%%%%%%%%%%%%%%%%%%%%%%%%%%%
\begin{figure}[htb]
\centering
\includegraphics[height=2.2in]{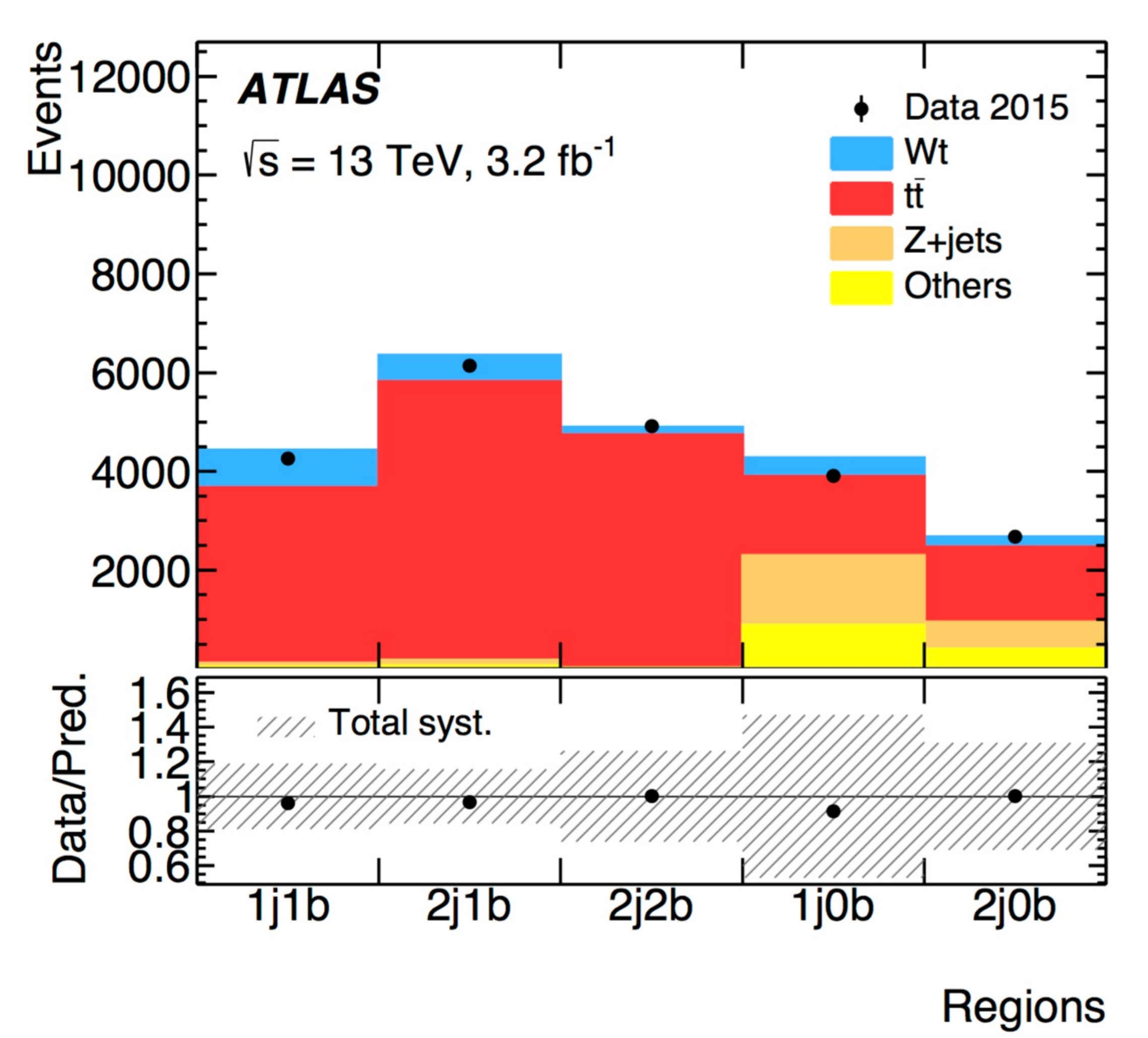}
\caption{Expected number of events for signal and backgrounds and observed number of data events are shown in the two signal regions (1j1b and 2j1b), the top-antitop control region (2j2b) and two additional validation regions (1j0b and 2j0b). The error bands represent the total systematic uncertainties \cite{Aaboud:2016lpj}.}
\label{fig:eventyields}
\end{figure}
%%%%%%%%%%%%%%%%%%%%%%%%%%%%%%%%%%%%%%%%%%%%%%%%%%%%%%%%%%%%%%%%%%%%%%%%%%%

Figure \ref{fig:eventyields} shows the expected number of events for signal and backgrounds as well as the observed data events in all the previously defined regions. The errors include the total systematic and statistical uncertainty. The signal and background samples are normalised to their respective theoretical predictions. The ratio between the observed and predicted event yields is also included and shows a very good agreement between the two. It is also evident that both signal regions are completely dominated by the top pair production background and that even with the dedicated selection, the signal to background ratio is small. 

In order to separate the signal from the top-antitop background, variables are combined into a single discriminant with increased separation power by using a boosted decision tree algorithm (BDT). Separate BDTs are trained for each of the two signal regions. Variables corresponding to different kinematic properties of the selected physics objects are given as input. For the 2j1b signal region, the variables with the largest separation power are chosen to be the transverse momenta of the all the objects in the final state, as well as the $p_T$ difference between the two lepton system and the jet+$E_T^{\text{miss}}$ one. In the 2j2b region the leading variable is the transverse momentum of the two lepton system, followed by the $\Delta R$ between the and the jet and missing transverse momentum system. 

\begin{figure}[htb]
 	\centering
 	\includegraphics[height=2in]{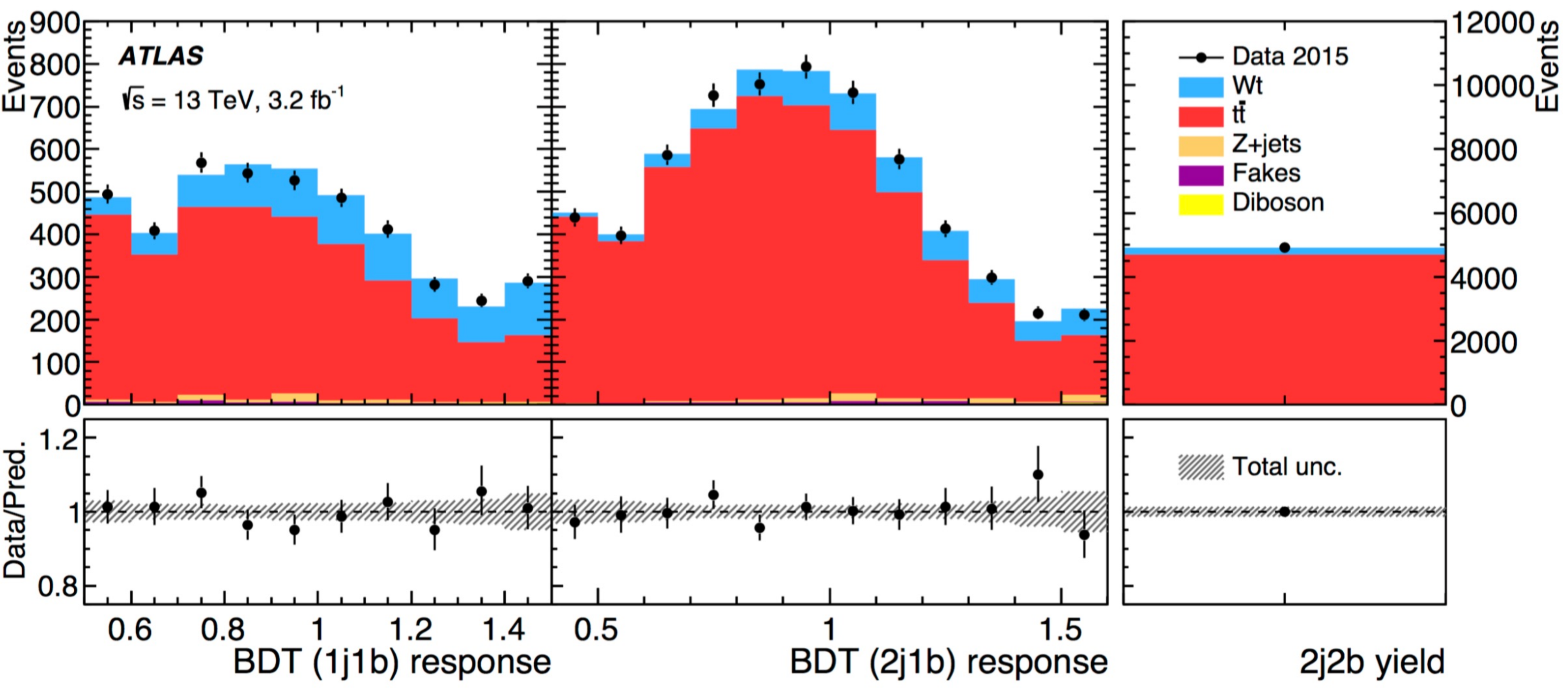}
 	\caption{ Distributions used for the binned-likelihood fit. In the signal regions (1j1b and 2j1b) the BDT response is used while in the 2j2b control region only the total number of events is considered \cite{Aaboud:2016lpj}.}
 	\label{fig:bdt}
 \end{figure}

In order to extract the $Wt$ cross-section, a binned likelihood fit is performed over the two signal regions (1j1b and 2j1b) and the 2j2b control region.  For the signal regions, the BDT discriminant is used in the fit, while in the control region, only the normalization is taken into account. The fitted templates are shown in Figure \ref{fig:bdt}.

Many sources of experimental systematic uncertainties are taken into account. These include the luminosity measurement, lepton efficiency scale factors used to correct simulation to data, lepton energy scale and resolution, missing transverse energy related terms and the efficiency of identifying jets coming from $b$-quarks. The dominant systematic for this analysis is related to the measurement of the jet energy scale and resolution. 
 In addition to the experimental systematics, uncertainties that arise due to theoretical modelling for the signal and $t\bar{t}$ background are also evaluated. The dominant uncertainty in this category is the NLO matrix element generator and the parton shower and hadronisation generator. 
 
 The total systematics uncertainty is 30\% on the measured $Wt$ cross-section. 
 %%%%%%%%%%%%%%%%%%%%%%%%%%%%%%%%%%%%%%%%%%%%%%%%%%%%%%%%%%%%%%%%%%%%%%%%%
 %%%%%%%%%%%%%%%%%%%%%%%%%%%%%%%%%%%%%%%%%%%%%%%%%%%%%%%%%%%%%%%%%%%%%%%%%%%

%%%%%%%%%%%%%%%%%%%%%%%%%%%%%%%%%%%%%%%%%%%%%%%%%%%%%%%%%%%%%%%%%%%%%%%%%
%%
%%   use this format to include a LaTeX table  into your paper
%%
%\begin{table}[t]
%\begin{center}
%\begin{tabular}{l|ccc}  
%Patient &  Initial level($\mu$g/cc) &  w. Magnet &  
%w. Magnet and Sound \\ \hline
% Guglielmo B.  &   0.12     &     0.10      &     0.001  \\
% Ferrando di N. &  0.15     &     0.11      &  $< 0.0005$ \\ \hline
%\end{tabular}
%\caption{Blood cyanide levels for the two patients.}
%\label{tab:blood}
%\end{center}
%\end{table}
%%%%%%%%%%%%%%%%%%%%%%%%%%%%%%%%%%%%%%%%%%%%%%%%%%%%%%%%%%%%%%%%%%%%%%%%%%%

\section{Results}

An inclusive measurement of the $Wt$ associated production at 13 TeV is performed by fitting templates to the BDT discriminant distributions in three separate regions that include two leptons, missing transverse energy but differ in the number of jets and $b$-tagged jets. The cross-section is extracted to be $\sigma_{Wt} = 94\pm 10\text{(stat.)}^{+28}_{-22}\text{(syst.)}\pm 2\text{(lumi})$ pb. The result has an observed (expected) significance of 4.5 $\sigma$ (3.9 $\sigma$). This result is in good agreement with the Standard Model expectation calculated at next-to-leading order (NLO) with next-to-next-to-leading logarithmic (NNLL) soft-gluon corrections $\sigma_{\text{theory}} = 71.7\pm 1.8\text{(scale)}\pm 3.4(\text{PDF})$ pb \!\cite{Kidonakis:2015nna}.

\end{document}